\documentclass{aa}  
\usepackage{graphicx}
\usepackage{txfonts}

\usepackage{color}

\begin{document}
   \title{A probable close brown dwarf companion to GJ~1046 (M2.5V)
          \thanks{Based on observations collected at
          the European Southern Observatory, Paranal, Chile, programmes
          173.C-0606 and 078.C-0829}
          }


   \author{M. K\"urster,
          \inst{1}
          M. Endl
          \inst{2}
          and
          S. Reffert
          \inst{3}
          }

   \offprints{M. K\"urster}

   \institute{Max-Planck-Institut f\"ur Astronomie,
              K\"onigstuhl 17, D-69117 Heidelberg\\
              \email{kuerster@mpia-hd.mpg.de}
         \and
              McDonald Observatory, University of Texas, Austin, TX 78712\\
              \email{mike@astro.as.utexas.edu}
         \and
              Zentrum f\"ur Astronomie Heidelberg, Landessternwarte, 
              K\"onigstuhl 12, D-69117 Heidelberg\\
              \email{sreffert@lsw.uni-heidelberg.de}
             }

   \date{Received 18 January 2008 / Accepted 17 March 2008}

 
  \abstract
   {Brown dwarf companions to stars at separations of a few AU or less are
    rare objects, and none have been found so far around early-type M dwarfs
    (M0V-M5V). With GJ~1046 (M2.5V), a strong candidate for such a system 
    with a separation of $0.42~\mathrm{AU}$ is presented.}
   {We aim at constraining the mass of the companion in order to decide
    whether it is a brown dwarf or a low-mass star.}
   {We employed precision RV measurements to determine the
    orbital parameters and the minimum companion mass.
    We then derived an upper limit to the companion mass from the
    lack of disturbances of the RV measurements by a secondary spectrum.
    An even tighter upper limit is subsequently established by combining
    the RV-derived orbital parameters with the recent new version of 
    the Hipparcos Intermediate Astrometric Data.}
   {For the mass of the companion, we derive 
    $m \ge 26.9~\mathrm{M}_\mathrm{Jup}$ from the RV data. 
    Based on the RV data alone, the
    probability that the companion exceeds the stellar mass threshold is
    just $6.2\% $. 
    The absence of effects from the secondary spectrum lets us constrain
    the companion mass to $m \le 229~\mathrm{M}_\mathrm{Jup}$.
    The combination of RV and Hipparcos data yields a $3\sigma $
    upper mass limit to the companion mass of $112~\mathrm{M}_\mathrm{Jup}$ 
    with a formal optimum value at $m=47.2~\mathrm{M}_\mathrm{Jup}$.
    From the combination of RV and astrometric data, the chance probability 
    that the companion is a star is $2.9\% $.}
   {We have found a low-mass, close companion to an early-type M dwarf.
    While the most likely interpretation of this object is that it is a brown
    dwarf, a low-mass stellar companion is not fully excluded.}

   \keywords{Stars: low-mass, brown dwarfs -- Binaries: spectroscopic --
             Stars: individual: GJ1046 -- Astrometry
               }
   \authorrunning{M.~K\"urster et al.}
   \maketitle
%

\section{Introduction}

The paucity of brown dwarf companions to solar-like 
stars at separations of
a few AU or less (a canonical value of $\le 5\mathrm{AU}$ is usually quoted),
was already noted by Campbell et~al.~(1988) in their early precision
radial velocity survey. This ``brown dwarf desert'' is currently
not well understood (see Grether \& Lineweaver 2006 for an overview).
Two distinctive formation mechanisms seem
to be at work for planetary ($M\le 13~\mathrm{M}_\mathrm{Jup}$) and stellar 
($M\ge 0.08~\mathrm{M}_\odot $) companions with relatively little 
overlap between the two. At wide separations no ``brown dwarf desert'' is  
observed. While the frequency of brown dwarf companions separated from
their host star by $<3~\mathrm{AU}$ is about $0.5\% $, it is at least a 
factor of $10$ higher for separations $>1000~\mathrm{AU}$
(Gizis et~al.~2001; also Neuh\"auser \& Guenther 2004).
The fact that close-in brown dwarf companions are rare 
is highly significant since the commonly employed radial velocity (RV) 
method to search for sub-stellar companions to stars is very
sensitive to such objects. The RV semi-amplitude $K$ of the 
primary (the companion is usually not visible
in the spectrum) is given by 
\begin{equation}
K = {(2\pi G/P)^{1/3}\over (1-e^2)^{1/2}}~{m\sin i\over (M + m)^{2/3}}~,
\end{equation}
where
$M$ and $m$ are, respectively, the mass of the star and companion,
and $P$, $e$ and $i$ are the orbital period, eccentricity and inclination.

As can be seen from equation~(1) the chances to detect a companion object 
via RVs increase with shorter period (and shorter separation) and higher 
companion mass. For example, the RV semi-amplitude of the stellar reflex
motion caused by a $20~\mathrm{M}_\mathrm{Jup}$ brown dwarf at 
$1~\mathrm{AU}$ from a solar-mass star is $565~\mathrm{ms}^{-1}$, 
if the orbit is circular and seen edge-on ($i=90^\circ $). 
This is two orders of magnitude greater than the current state-of-the-art 
RV measurement precision of a few $\mathrm{ms}^{-1}$. 
The detectability also increases for lower-mass stars;
the same $20~\mathrm{M}_\mathrm{Jup}$ at $1~\mathrm{AU}$ from an
$0.3~\mathrm{M}_\odot $ star would produce an RV semi-amplitude of 
$1009~\mathrm{ms}^{-1}$.  And in fact there is observational evidence of 
the existence of brown dwarf companions to low-mass stars such as the
prototype brown dwarf companion GJ~229B that orbits an M1V star
at a wide projected separation of $44~\mathrm{AU}$ (Nakajima et~al. 1995).

   \begin{figure*}
   \includegraphics[width=13.0cm]{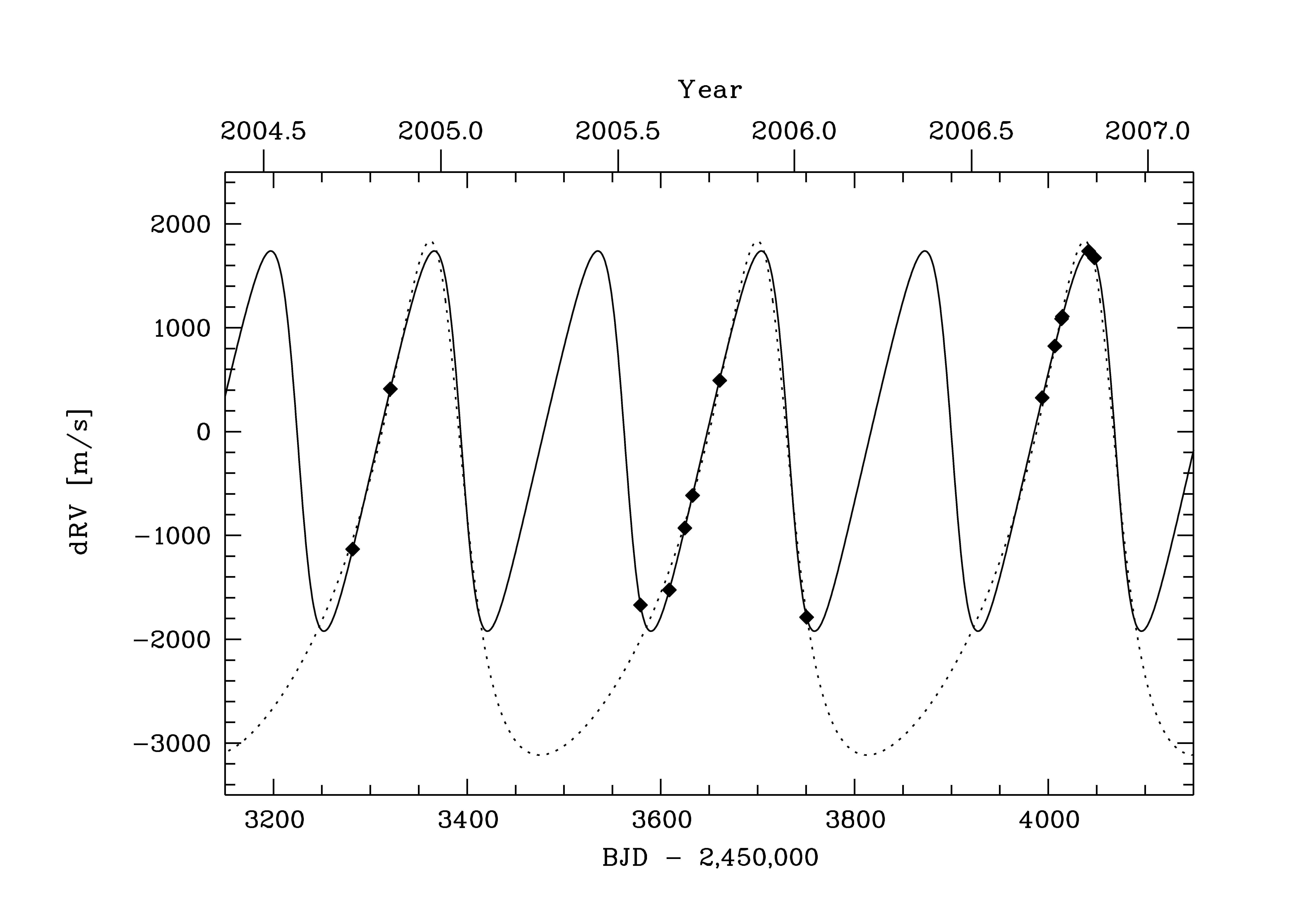}
      \caption{RV time series of our UVES RV data for GJ~1046.
               The solid line corresponds to the Keplerian solution with 
               a period of $169~\mathrm{d}$ and with $\chi ^2 = 12.7$, 
               8 degrees of freedom (DoF), and $p(\chi ^2)=0.123$.
               The rms scatter of the
               measurements around the orbital solution is 
               $3.56~\mathrm{ms}^{-1}$ which compares well with the
               average measurement error of $3.63~\mathrm{ms}^{-1}$
               which is much smaller than the plot symbols.
               For comparison the second best solution with a period of
               $338~\mathrm{d}$ is shown as a dashed line.
               While being the second best, the latter solution is clearly 
               excluded due to its extremely large $\chi ^2$ value of
               $10~121$.
              }
         \label{FigTimSer}
   \end{figure*}
%

Few brown dwarf companion candidates in the separation regime 
up to a few AU are known.
The first such candidate was HD~114762 (Latham et~al.~1989), and a more recent
example is HD~137510 (Endl et~al.~2004).
The masses of these candidate objects have often not been well-determined
since the RV method just yields a minimum mass and 
the astrometric precision of the available Hipparcos data (ESA 1997)
is mostly not sufficient to confirm brown dwarfs and exclude stellar
companions (e.g.~Pourbaix 2001; Pourbaix \& Arenou 2001). 
Among the best established brown dwarfs are the companions to the
G4IV star HD~38529 and to the G6IV star HD~168443 with companion masses of 
$37~\mathrm{M}_\mathrm{Jup}$ and $34~\mathrm{M}_\mathrm{Jup}$, 
orbital periods of $2174.3~\mathrm{d}$ and $1770~\mathrm{d}$, and 
separations of $3.68~\mathrm{AU}$ and $2.87~\mathrm{AU}$, respectively. 
These masses were determined by Reffert \& Quirrenbach (2006) who derived
new astrometric solutions from the Hipparcos measurements given the precisely 
known RV-derived orbital parameters, i.e.~period, time of periastron passage, 
eccentricity, and longitude of periastron. Another example determined in a 
similar fashion by Zucker \& Mazeh (2000) is the G5IV star 
HD~10697 with a companion in a $1078~\mathrm{d}$ orbit at a separation of
$2.12~\mathrm{AU}$ and a mass of $40~\mathrm{M}_\mathrm{Jup}$.
An example for an object with a minimum mass of $9.3~\mathrm{M}_\mathrm{Jup}$
that turned out to be a star with a mass of $142~\mathrm{M}_\mathrm{Jup}$
is the companion to HD~33636 (Bean et~al.~(2007).

The present paper presents a new candidate for a brown-dwarf companion 
which we have found in our precision RV survey carried out with the UVES 
spectrograph at the ESO VLT in search for planetary and substellar companions
to M dwarfs (see K\"urster et~al.~2003). The host star, 
GJ~1046 (M2.5V; $V=11.62~\mathrm{mag})$, has no entry in the
Double and Multiple Systems Annex of the Hipparcos data base.
Comparing its V band and J,H,K band colours (from the 2MASS catalogue; 
Skrutskie et~al.~2006) with the mass-luminosity relationships
by Delfosse et~al.~(2000) there is no indication of near-infrared
emission in excess of the scatter found in these relations.

If the companion to GJ~1046 turns out to be a brown dwarf,
then the system would be unique in that it would contain the first close-in
$(\le 5~\mathrm{AU}$)
brown dwarf companion to a main-sequence star of spectral type
early-M (M0V--M5V). Since the mass ratios of binary systems with low-mass
primaries tend towards 
unity, brown dwarf companions to late-M dwarfs ($\ge$M6V) are relatively
frequent (e.g.~Montagnier et~al.~2006). But even in these systems
separations $\le 5~\mathrm{AU}$ are usually found only among binaries with 
low-mass stellar secondaries. Counterexamples are the M8 star LHS 2397a with 
an L7.5 companion at a separation of $2.9~\mathrm{AU}$ (Freed et al.~2003)
and the young M6 object at the star-to-brown dwarf border, Cha~H$\alpha 
$~8, orbited by a brown dwarf at a separation of $1\mathrm{AU}$
(Joergens \& M\"uller 2007).

\begin{table}
\caption{Differential RV time series measurements of GJ~1046. 
}
\label{table:1}
\centering
\begin{tabular}{ccrc}
\hline\hline
 Date & $^\mathrm{a)}~\mathrm{BJD}~-$ & DRV~~ & RV-error \\
 & $2~450~000$ & $[\mathrm{ms^{-1}}]$~ & $[\mathrm{ms^{-1}}]$ \\
\hline
2004-10-03 & 3281.83304 & -1132.0 & 3.2 \\
2004-11-11 & 3320.62631 &   411.5 & 3.5 \\
2005-07-27 & 3578.91097 & -1670.6 & 4.8 \\
2005-08-26 & 3608.77350 & -1524.6 & 3.7 \\
2005-09-11 & 3624.71807 &  -928.7 & 4.3 \\
2005-09-19 & 3632.72200 &  -615.2 & 3.5 \\
2005-10-17 & 3660.68277 &   492.3 & 3.5 \\
2006-01-15 & 3750.60256 & -1787.7 & 3.5 \\
2006-09-15 & 3993.84856 &   326.4 & 3.3 \\
2006-09-28 & 4006.77107 &   823.4 & 3.9 \\
2006-10-05 & 4013.80030 &  1085.8 & 3.6 \\
2006-10-06 & 4014.61077 &  1108.1 & 3.7 \\
2006-11-02 & 4041.65276 &  1737.5 & 3.1 \\
2006-11-08 & 4047.65399 &  1673.7 & 3.3 \\
\hline
\end{tabular}
\begin{tabular}{ll}
    ~ & ~ \\
Note: & a) Barycentrically corrected Julian Date ~ \\
\end{tabular}
\end{table}

\section{Observations}

GJ~1046 was observed with the VLT-UT2+UVES as one of the targets of
our precision RV survey of M dwarfs in search for extrasolar planets
(see K\"urster et~al.~2003, 2006). 
To attain high-precision RV measurements UVES was self-calibrated
with its iodine gas absorption cell operated at a temperature
of $70^{\circ }\mathrm{C}$. 
Image slicer \#3 and an $0.3\arcsec $ slit were chosen
yielding a resolving power of $R = 100~000-120~000$. The central wavelength 
of $600~\mathrm{nm}$ was selected such that the useful spectral
range containing iodine ($\mathrm{I}_2$) absorption lines 
$(500-600~\mathrm{nm})$
falls entirely on the better quality CCD of the mosaic of two
4~K$~\times ~2$~K CCDs.

\begin{table}
\caption{System parameters of GJ~1046}
\label{table:2}
\centering
\begin{tabular}{lccc}
\hline\hline
\multicolumn{4}{l}{\bf RV-derived parameters} \\
\hline
 & & & \\
Orbital period $P$ & $168.848$ &$\pm 0.030$ & $[\mathrm{d}]$ \\
Time of periastron $T_\mathrm{p}$ & & & \\
$~~~~~~~BJD-2~450~000$ & $3~225.78$ & $\pm 0.32$ & \\
RV semi-amplitude $K$ & $1830.7$ & $\pm 2.2$ & $[\mathrm{ms}^{-1}]$ \\
Orbital eccentricity $e$ & $0.2792$ & $\pm 0.0015$ & \\
Longitude of periastron $\omega $ & $92.70$ & $\pm 0.50$ & $[^\circ ]$ \\
Mass function $f(m)$ & $9.504$ & $\pm 0.024$ & $[10^{-5}\mathrm{M}_\odot ]$ \\
$\chi ^2$ (DoF$=8$)          & 12.7 &  & \\
$p(\chi ^2)$        & 0.123 & & \\
Scatter rms & 3.56 & & $[\mathrm{ms}^{-1}]$ \\
Mean error $\overline{\Delta RV}$ & 3.63 & & $[\mathrm{ms}^{-1}]$ \\
 & & & \\
\hline\hline
\multicolumn{4}{l}{\bf Inferred parameters} \\
\hline
 & & & \\
Stellar mass $M$ & $0.398$ & $\pm 0.007$ & $[\mathrm{M}_\odot ]$ \\
Minimum companion & & & \\
~~~~~~~mass $m_\mathrm{min}$ & $26.85$ & $\pm 0.30$ & $[\mathrm{M}_\mathrm{Jup}]$ \\
Semi-major axis of & & & \\
~~~~~~~companion orbit $a$ & $0.421$ & $\pm 0.010$ & $[\mathrm{AU}]$ \\
Critical inclination & & & \\
~~~~~~~(for $m=0.08\mathrm{M}_\odot $) $i_\mathrm{crit}$ & $20.4$ & & $[^\circ ]$ \\
$^\mathrm{a)}$ Probability of $i_\mathrm{crit}$, ~ $p_{i_\mathrm{crit}}$ & $6.3\% $ & \multicolumn{2}{l}{$(20.4^\circ >i\ge 0^\circ )$} \\
 & & & \\
\hline\hline
\multicolumn{4}{l}{\bf Parameters derived from absence of companion spectrum}
 \\
\hline
 & & & \\
Maximum companion & & & \\
~~~~~~~mass $m_\mathrm{max}^\mathrm{~sp}$ & $229$ & & $[\mathrm{M}_\mathrm{Jup}]$ \\
Minimum inclination $i_\mathrm{min}^\mathrm{~sp}$ & $8.7$ & & $[^\circ ]$ \\
$^\mathrm{a)}$ Probability of $i_\mathrm{min}^\mathrm{~sp}$, ~ $p_{i_\mathrm{min}^\mathrm{~sp}}$ & $1.2\% $ & \multicolumn{2}{l}{$(8.7^\circ >i\ge 0^\circ )$} \\
 & & & \\
\hline\hline
\multicolumn{4}{l}{\bf Astrometry-derived parameters} \\
\hline
 & & & \\
Ascending node $\Omega $ & $97.7$ & {\scriptsize formal optimum} & $[^\circ ]$ \\
Inclination $i$ & $125.9$ & {\scriptsize formal optimum} & $[^\circ ]$ \\
Companion mass $m$ & $47.2$ & {\scriptsize formal optimum} & $[\mathrm{M}_\mathrm{Jup}]$ \\
$\chi ^2$ (DoF$=202$) & $329.0$ &  & \\
Minimum inclination $i_\mathrm{min}^\mathrm{~as}$ & $15.6$ & $3\sigma $ limit & $[^\circ ]$ \\
$^\mathrm{a)}$ Probability of $i_\mathrm{min}^\mathrm{~as}$, ~ $p_{i_\mathrm{min}^\mathrm{~as}}$ & $3.7\% $ & \multicolumn{2}{l}{$(15.6^\circ >i\ge 0^\circ )$} \\
Maximum companion & & & \\
~~~~~~~mass $m_\mathrm{max}^\mathrm{as~}$ & $112$ & $3\sigma $ limit & $[\mathrm{M}_\mathrm{Jup}]$ \\
$^\mathrm{b)}$ Probability of a stellar & & & \\
~~~~~~~ companion $p_\ast $ & $2.9\% $ & & \\
 & & & \\
\hline
\end{tabular}
\begin{tabular}{ll}
    ~ & ~ \\
Notes: & $\mathrm{a)}$ {\em A priori} probability based on the RV derived 
         minimum \\
       &           ~~~ mass and assuming random orientation of the orbit. \\
       & $\mathrm{b)}$ Probability derived from the astrometric model. \\
\end{tabular}
\end{table}


For a detailed description of our data modelling approach employed for the 
determination of high precision differential radial velocities $(DRV)$ we 
refer the reader to Endl et al. (2000). A concise summary can also be found 
in Sect.~4 of K\"urster et~al.~(2003).

A total of 14 spectra of GJ~1046 observed through the iodine cell were 
obtained in 14 nights between 03 October 2004 and 11 November 2006. See
Table~1 for the journal of observations.
Individual exposure time was $900~\mathrm{s}$ yielding an average S/N per
pixel between 39 and 58 for the various spectra (the median and mean being
$51.0$ and $49.1$, respectively). 
On average our error of the individual RV measurements is 
$3.63~\mathrm{ms}^{-1}$ for this $V=11.62~\mathrm{mag}$ object.
All RV data were corrected to the solar system barycenter using the
JPL ephemeris DE200 (Standish 1990) for the flux-weighted
temporal midpoint of the exposure as given by the UVES exposuremeter.
For each epoch of observation proper motion corrected
stellar coordinates were used. On 19 November 2004 we also
obtained a triplet of exposures without the iodine cell 
(exposure time $3\times 705~\mathrm{s}$) required as a template spectrum
in the data modelling process (cf.~Endl et~al.~2000).

\section{Results}

Our RV time series for GJ~1046 is listed in Table~1 and also shown in 
Fig.~1 along with the best-fit Keplerian orbit yielding an orbital period 
of $P=169~\mathrm{d}$, an eccentricity of $e=0.28$, an 
RV semi-amplitude of $K=1831~\mathrm{ms}^{-1}$, and a mass function
$f(m) = (m\sin i)^3 / (M+m)^2 = 9.5\times 
10^{-5}~\mathrm{M}_\odot $ (Table~2).



   \begin{figure}
   \centering
   \includegraphics[width=8.8cm]{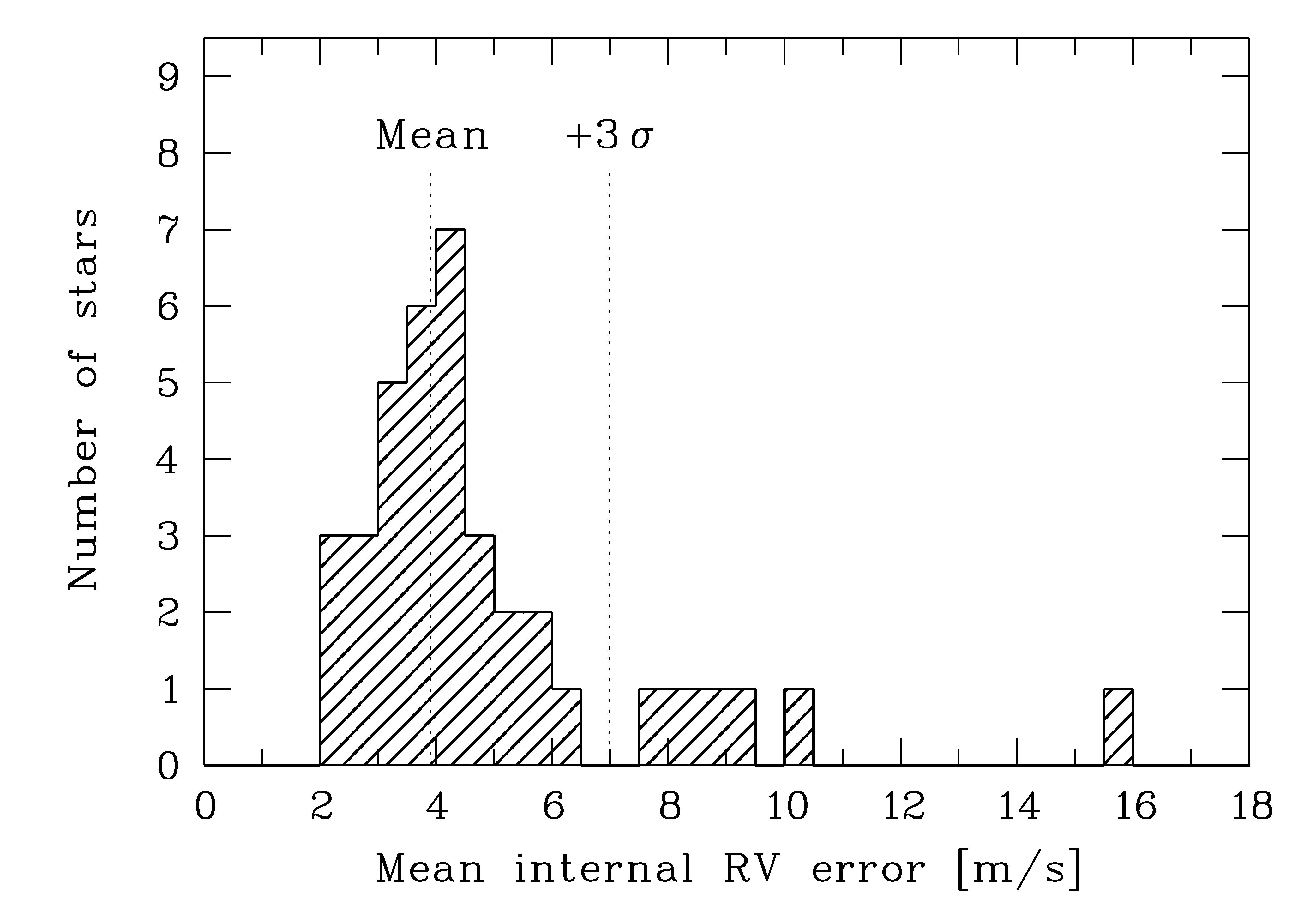}
      \caption{Histogram of the mean internal RV measurement errors for
               our sample of stars. Applying a $\kappa$-$\sigma $
               clipping procedure with an iterative rejection of values
               exceeding the mean plus $2.6~\sigma $ (for a Gaussian 
               distribution the chance probability of exceeding this value
               in one iteration is $0.5\% $) we reject all stars with
               mean RV errors in excess of $6.5~\mathrm{ms}^{-1}$.
               Of the total sample of 41 stars only 38 are displayed,
               because three stars with large errors 
               ($79$, $81$, and $92~\mathrm{ms}^{-1}$, respectively) 
               lie far outside the displayed error range. They are
               newly discovered double-lined spectroscopic binaries
               with such strong contributions from the secondaries
               that the employed single-lined spectrum model fails.
               Six stars have abnormally high mean errors between $7$ and 
               $16~\mathrm{ms}^{-1}$ which, in one case, is the result of
               very low signal-to-noise ratio, but in the other cases
               could also be due to spectral contamination from a companion.
               The remaining bulk of the
               distribution (mean error $<6.5~\mathrm{ms}^{-1}$) has a
               mean of $3.91~\mathrm{ms}^{-1}$ and a width of
               $\sigma = 1.02~\mathrm{ms}^{-1}$.
              }
         \label{FigChi3}
   \end{figure}
%


\begin{table*}
\caption{Simulations for the determination of an upper limit to the companion
mass.
}
\label{table:3}
\centering
\begin{tabular}{cccccclll}
\hline\hline
\multicolumn{2}{c}{Mass} & V-band & Model & Mean error & Excess error 
            & \multicolumn{3}{c}{Chance probability from} \\ 
$[\mathrm{M}_\odot ]$ & $[\mathrm{M}_\mathrm{Jup} ]$ &
            flux ratio & spectrum & $[\mathrm{ms}^{-1}]$ & 
            $[\mathrm{ms}^{-1}]$ & \multicolumn{2}{c}{sample comparison} & 
            excess error \\
 & & & & & & ~~~a) & ~~~b) & criterion \\
\hline
0.00 & ~~0 & ---   & none   & 3.63 & 0.00 & 0.56  & 0.61    & 1    \\
\hline                                                                    
0.16 & 167 & 18.30 & GJ~699 & 3.79 & 1.07 & 0.50  & 0.55    & 0.70 \\
0.17 & 177 & 14.85 & GJ~699 & 3.92 & 1.49 & 0.47  & 0.50    & 0.59 \\
0.18 & 188 & 12.26 & GJ~699 & 4.04 & 1.77 & 0.47  & 0.45    & 0.51 \\
0.19 & 198 & 10.27 & GJ~699 & 4.22 & 2.16 & 0.34  & 0.38    & 0.40 \\
0.20 & 209 & ~8.69 & GJ~699 & 4.47 & 2.61 & 0.25  & 0.29    & 0.28 \\
0.20 & 209 & ~8.69 & GJ~682 & 4.41 & 2.51 & 0.31  & 0.31    & 0.31 \\
0.21 & 219 & ~7.43 & GJ~682 & 4.79 & 3.12 & 0.17  & 0.20    & 0.14 \\
0.22 & 230 & ~6.41 & GJ~682 & 5.16 & 3.67 & 0.13  & 0.11    & 0    \\
0.23 & 240 & ~5.57 & GJ~682 & 5.61 & 4.28 & 0.094 & 0.048   & 0    \\
0.24 & 250 & ~4.87 & GJ~682 & 6.10 & 4.90 & 0.031 & 0.016   & 0    \\
0.25 & 261 & ~4.29 & GJ~682 & 6.70 & 5.63 & 0     & 0.0032  & 0    \\
0.26 & 271 & ~3.79 & GJ~682 & 7.40 & 6.45 & 0     & 0.00032 & 0    \\
\hline
\end{tabular}
\begin{tabular}{ll}
    ~ & ~ \\
Notes: & a) Fraction of stars in the distribution with larger mean errors \\
       & b) Areal fraction of Gaussian fitted to the distribution 
            exceeding the mean error ~~~~~~~~~~~~~~~~~~~~~~~~~~~~ \\
\end{tabular}\end{table*}

Given the unknown inclination $i$ only a minimum to the companion mass 
can be determined corresponding to the case $i=90^\circ $.
For this we need an estimate of the stellar mass of 
GJ~1046 which we obtain from the K-band 
mass-luminosity relationship by Delfosse et~al. (2000). 
Taking the apparent K-magnitude  
from the 2MASS catalogue $(K=7.03~\mathrm{mag})$ and combining it with the 
Hipparcos parallax $(71.11~\mathrm{mas})$ we find an absolute 
K-magnitude of $6.29~\mathrm{mag}$. The K-band mass-luminosity relationship
then yields a stellar mass of $0.398\pm 0.007~\mathrm{M}_\odot $. 

We then infer a minimum companion mass of 
$m_\mathrm{min} = 26.9~\mathrm{M}_\mathrm{Jup}$ 
and, from equation~(1), a semi-major axis of the companion orbit of 
$a=0.42~\mathrm{AU}$. 
In order for the true companion mass to exceed the stellar threshold of 
$0.08~\mathrm{M}_\odot $ the orbital inclination $i$ would have to be 
$<20.4^\circ $. For a chance orientation of the orbit the 
probability that $i$ is smaller than some angle 
$\theta $ is given by 
\begin{equation}
p(\theta > i \ge 0^\circ )=1-\cos\theta ~;
\end{equation}
hence the chance probability to have an inclination $<20.3^\circ $ 
is just 6.3\% making it not very likely that the companion is a star
(see also Table~2).


\section{Spectroscopic companion mass upper limit}

An upper limit to the mass of the companion can be determined from the
spectroscopic data by exploiting the notion that with increasing mass the 
companion would at some point become so bright that it would noticeably 
affect the RV measurements. In our data modelling approach a suitable 
indicator for the presence of an additional perturbing signal is the 
magnitude of the RV measurement error, because all RV data are obtained 
assuming that the modelled spectrum is that of a single-lined binary  
in which the light from the companion can be neglected. 
If this is not the case, then the light contribution
from the companion manifests itself as unusually large errors of the 
determined RV values. For details of the data modelling approach and the
estimation of the RV errors we refer the reader to Endl et~al.~(2000).

Briefly, the spectra are subdivided into $\approx 500$ chunks of
$\approx 2~\AA $ width, each of which is modelled in order to yield an
independent RV measurement. The mean of these values and its error are taken
as the final measurements and their (internal) error estimates.
If a second contaminating spectrum is present, the different spectral chunks
are affected in a non-homogeneous fashion, depending on the detailed
spectral line patterns within the chunk, with the effect that the derived 
chunk RV values exhibit a stronger scatter and hence combine to
an increased value of the internal RV measurement error. 
By way of simulations adding faint companion spectra to the
spectra of GJ~1046 we investigate the following two possibilities of obtaining 
information on the companion mass upper limit from the internal RV error.

\begin{enumerate}
   \item A comparison of the mean internal RV measurement error of GJ~1046
         for different added companion spectra with the distribution of mean
         observed errors for the sample of monitored stars.
   \item A comparison of the mean internal RV measurement error of GJ~1046
         for different added companion spectra with its original value.
\end{enumerate}

In the case of considerable contamination the average RV error
for the spectra of the star in question stands out from the 
distribution of RV errors for the sample of monitored stars which is shown in
Fig.~2. If only the bulk of the distribution is considered, i.e.~stars with 
a mean RV error $<7~\mathrm{ms}^{-1}$, this distribution is nearly Gaussian 
with a mean value of
$3.91~\mathrm{ms}^{-1}$ and a width of $\sigma = 1.02~\mathrm{ms}^{-1}$.

The average RV measurement error for GJ~1046 is $3.63~\mathrm{ms}^{-1}$
with the errors of the 14 individual GJ~1046 spectra ranging from  
$3.07$ to $4.84~\mathrm{ms}^{-1}$. These values are typical of the range of
signal-to-noise values of our spectra (between $39$ and $58$ for GJ~1046)
and well inside the distribution of errors in our sample of stars (Fig.~3).
We also note that the RV errors increase with 
decreasing signal-to-noise ratio of the GJ~1046 spectra which would not be 
the case, if the errors were dominated by a perturbing signal.

In the performed simulations we added spectra of faint low-mass 
stars to the original GJ~1046 spectra. As summarised in Table~3 these
simulations explored the companion mass regime $0.16-0.26~\mathrm{M}_\odot $
(first two table columns) which, according to the V-band mass-luminosity 
relation by Delfosse et~al.~(2000), corresponds to companions that are 
factors of $18-4$ fainter than the stellar primary (third column). 
We have used the high-signal-to-noise template
spectra (observed without the iodine cell) of two other M dwarfs of 
this survey, GJ~699 (Barnard's star; M4V; mean S/N per pixel: 300) and 
GJ~682 (M3.5V; mean S/N per pixel: 230) for 
masses $\le 0.20~\mathrm{M}_\odot $ and $\ge 0.20~\mathrm{M}_\odot $, 
respectively (fourth column). For each simulated spectrum the companion 
spectrum was
shifted to the appropriate companion RV for the probed mass value
as well as scaled in flux corresponding to the brightness predicted
by the V-band mass-luminosity relation.

We find that a companion with a mass $\ge 0.254~\mathrm{M}_\odot $ or
$265~\mathrm{M}_\mathrm{Jup}$ and a factor of $4.1$ fainter in the V-band 
than its primary would increase the internal error (fifth column in Table~3) 
to $\ge 7~\mathrm{ms}^{-1}$ which would make this star stick out from 
the bulk of the sample (Fig.~2) indicating a contamination of the 
spectrum.
The 7th and 8th columns of Table~3 list the chance probabilities
of the obtained mean errors (fifth column) from a comparison
with the total sample of observed stars.

   \begin{figure*}
   \sidecaption
   \centering
   \includegraphics[width=8.8cm,bb= -30 0 410 400]{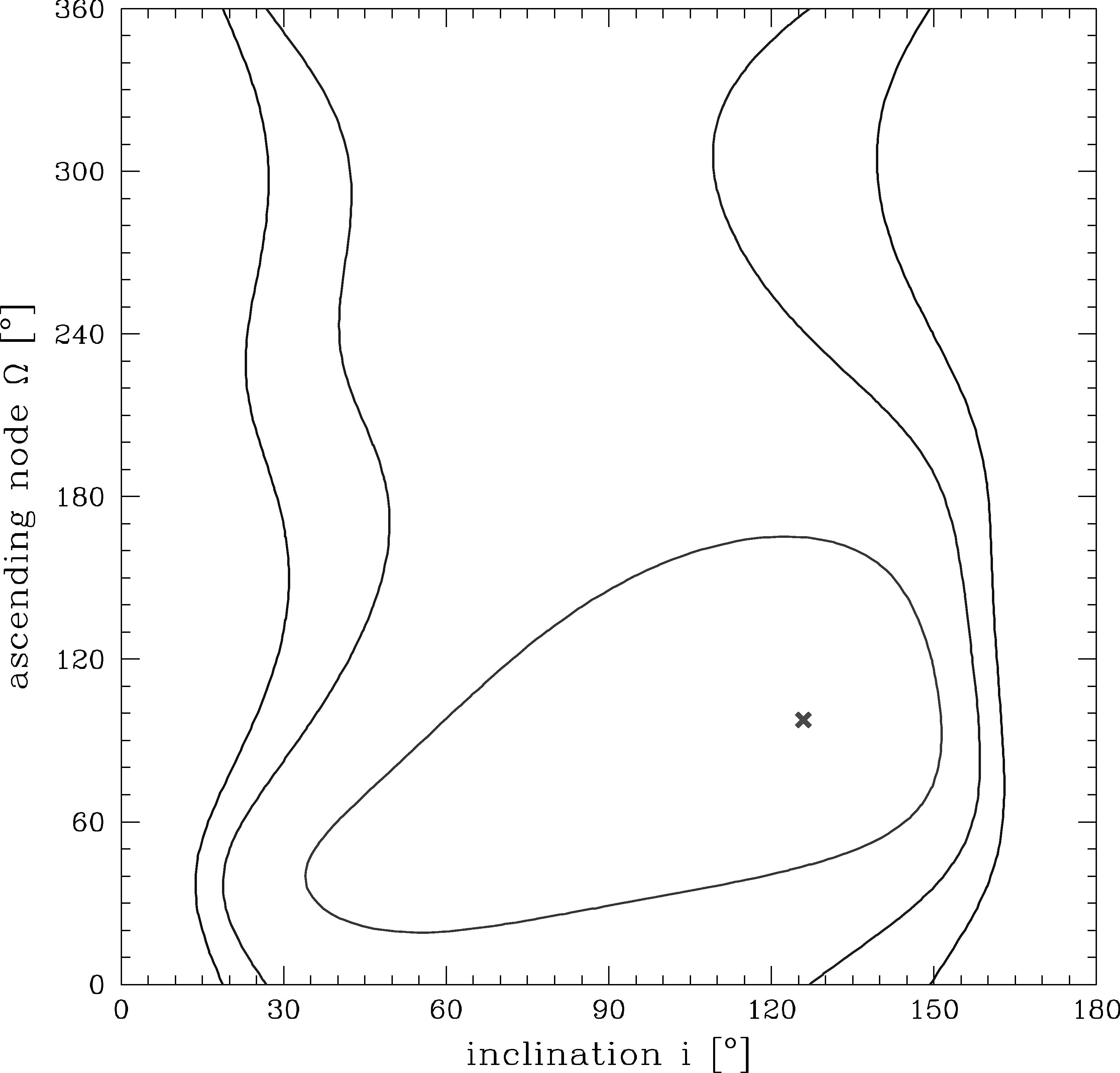}
   \includegraphics[width=8.8cm,bb= -40 0 400 400]{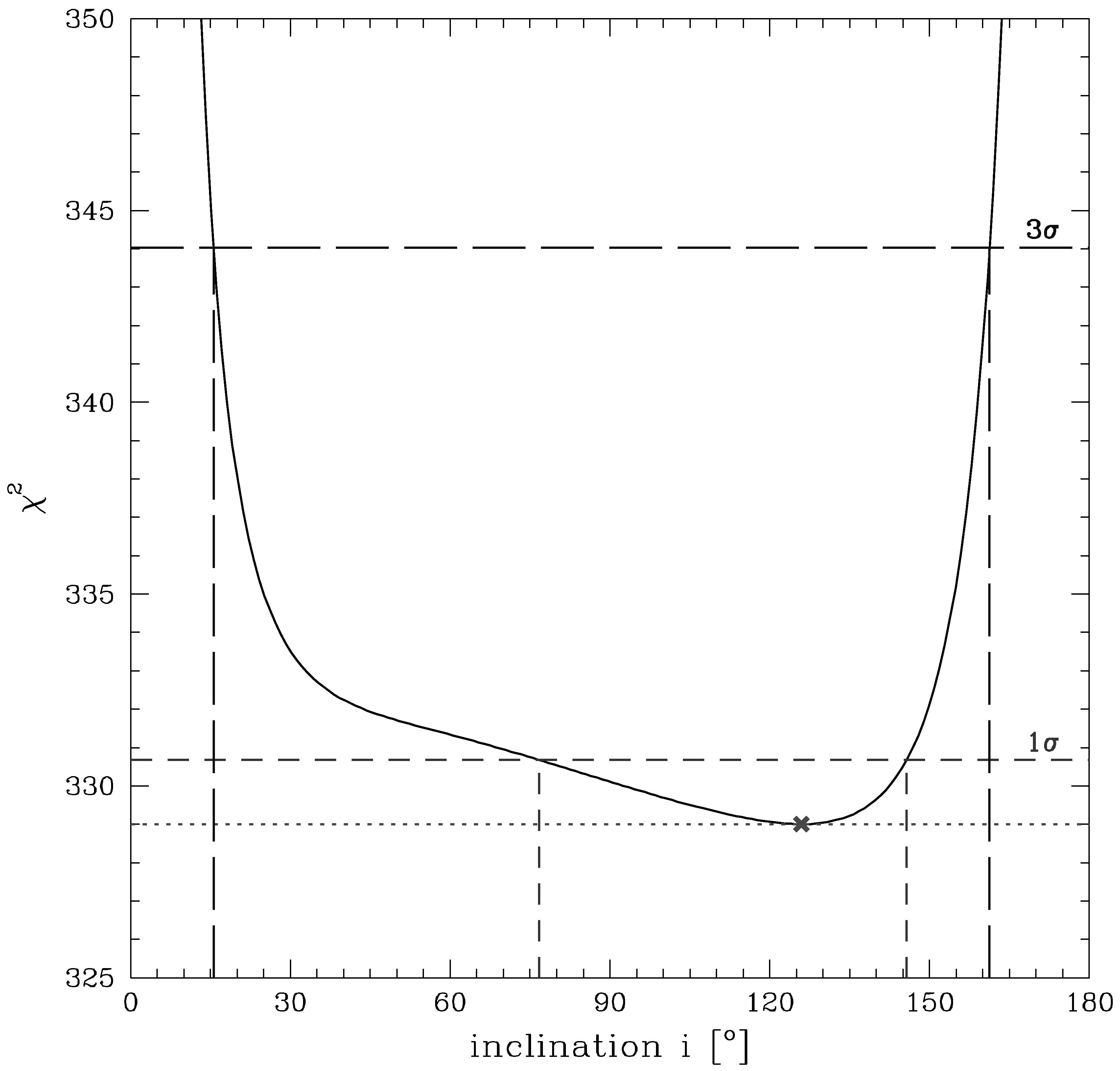}
      \caption{{\bf Left:} $\chi ^2$ contours for fitting a substellar 
               companion with
               fixed spectroscopic parameters to the Hipparcos Intermediate
               Astrometric Data of GJ~1046. The inclination $i$ and the 
               ascending node $\Omega $ were free parameters of the fit, 
               as were corrections to the standard five astrometric
               parameters in the Hipparcos Catalogue. The contours
               represent two-parameter joint confidence levels 
               with probabilities of $68.3\% $ $(1\sigma )$, $95.4\% $
               $(2\sigma )$, and $99.7\% $ $(3\sigma )$.
               The best fit solution is indicated by a cross.
               {\bf Right:} The $\chi ^2$ of the astrometric orbit as a
               function of inclination only. In this case the $1\sigma $
               and $3\sigma $ confidence levels indicated by the horizontal
               dashed lines correspond only to the single parameter $i$
               treating $\Omega $ as an uninteresting parameter.
               Again the best fit solution is indicated by a cross.
               }
         \label{FigChi2}
   \end{figure*}
%


For the second possibility of determining the mass upper limit of the 
companion
we assume (conservatively) that the mean original RV error is entirely
caused by contributions from the companion spectrum and not attributable
to photon noise or to effects of instrumental nature or intrinsic to the star.
We then search for the companion spectrum (as a function of companion
mass and brightness) whose addition to the observed spectra introduces
an additional RV error (6th column in Table~3)
of the same magnitude, i.e.~it doubles the square
of the errors. With an original value of $3.63~\mathrm{ms}^{-1}$ we
search in the simulated data for the companion mass and brightness that
leads to a mean intrinsic error a factor of 
$\sqrt{2}$ larger, i.e.~$5.13~\mathrm{ms}^{-1}$.
(We note in passing that this value is in the $88.4\% $ 
percentile of the distribution of the stellar sample truncated at 
$7~\mathrm{ms}^{-1}$; see Fig.~2.)
The 9th column of Table~3 lists the chance probability
of obtaining the excess error value listed in the 6th column.

For this increased error value we find a companion mass of 
$\ge 0.219~\mathrm{M}_\odot $ or $229~\mathrm{M}_\mathrm{Jup}$ and a 
primary-to-secondary V-band flux ratio of $6.5$ (cf.~Table~3).

As the mass value derived with the criterion to double the square of the
mean internal error is lower than the one derived from the comparison with 
the star sample, we will adopt the value of $229~\mathrm{M}_\mathrm{Jup}$
as the spectroscopic upper limit to the mass of the companion to GJ~1046.
This value corresponds to an orbital inclination of $8.7^\circ $.
The probability for an inclination as small as (or smaller than) this
value is $1.2\% $, again assuming random orientation of the orbit
(see also Table~2).

\section{Companion mass upper limit from a combination of the RV data with 
Hipparcos measurements}

Even if the astrometric signature of the companion is not seen in the Hipparcos
data, Hipparcos astrometry can yield stringent upper mass limits on companions
detected via the radial velocity method.

Using the Hipparcos parallax $(71.11~\mathrm{mas})$ together with the
orbital parameters derived from the RV measurements we can predict the minimum 
astrometric signal of the stellar reflex motion to be 
$3.7~\mathrm{mas}$ peak-to-peak. This corresponds to the full minor axis
of the orbit.
Since the Keplerian fit to the RV data only permits the determination of 
the projected orbit of the stellar reflex motion, the true astrometric 
effect could be considerably higher. For the limiting inclination of 
$20.4^\circ $ the full minor axis of the stellar orbit would extend 
$10.6~\mathrm{mas}$ on the sky. 

We have analysed the Hipparcos Intermediate 
Astrometric Data for GJ~1046 (HIP~10812) using the new reduction of the raw
data (van Leeuwen 2007a,b).
We followed the approach described in Reffert \& Quirrenbach (2006) by
keeping those of the orbital parameters that are known from the analysis of
the RVs fixed and varying only the inclination and the ascending node while 
fitting an astrometric orbit to the abscissa residuals.
Additional free parameters in the fit were a correction to the
mean position, mean proper motion and parallax of the star. The result is 
shown in Fig.~3 (left panel). 

The formally best fit to the Hipparcos data is achieved 
with an inclination $i=125.9^\circ ~(i-90^\circ = 35.9^\circ )$ 
corresponding to a true companion mass of $47.2~\mathrm{M}_\mathrm{Jup}$ 
pointing at a brown dwarf companion (Table~2).
\footnote{Varying the RV derived parameters within their errors leads to
minute changes in the formal best-fit solution indicating that the
uncertainties of the latter are absolutely dominated by the astrometric data.}
However, an F-test measuring the variance improvement yields a probability of
$17\% $ for the detection of the astrometric orbit implying that is has not
been detected with significance. This can also be seen in Fig.~3 (left panel),
where the ascending node is completely undetermined since the $2$ and 
$3\sigma $ confidence contour levels span the entire parameter range.

In the right panel of Fig.~3, the $\chi ^2$ value is shown as a function of 
inclination only, 
together with the $1\sigma $ and $3\sigma $ confidence regions for the 
inclination. The $3\sigma $ ($99.73\% $ confidence) lower limit to the 
inclination is $i=15.6^\circ $ implying a $3\sigma $ upper mass limit for 
the companion of $112~\mathrm{M}_\mathrm{Jup}$.

Therefore, a stellar companion cannot be fully excluded, even though it is
unlikely. From the astrometric solution the chance probability for the
companion to have a stellar mass, or equivalently, for its inclination to 
be either $i<20.3^\circ $ or $>159.7^\circ $
is $2.2\% $ and $0.7\% $, respectively, corresponding to a combined chance
probability of $2.9\% $ (see also Table~2).
\footnote{The combined chance probability is given by one minus the product
of the confidences: $1-(1-2.2\% )(1-0.7\% )=2.88\% $.}


\section{Conclusions}

We have presented the discovery of a probable brown dwarf companion
to an M dwarf with an orbital period of just under 1/2 year and
a star-companion separation of $0.42~\mathrm{AU}$.
Our RV measurements provide a lower limit to the true
companion mass of $26.9~\mathrm{M}_\mathrm{Jup}$
and a chance probability of just
$6.2\% $ that the companion is actually a star.
From the absence of any indications of a secondary spectrum in our data
we can place an upper limit to the companion mass of 
$m=229~\mathrm{M}_\mathrm{Jup}$.

Combining our RV measurements with the 
Hipparcos Intermediate Astrometric Data from the 
recent new reduction by van~Leeuwen (2007a,b)
we find a formal best-fit companion mass value of
$47.2~\mathrm{M}_\mathrm{Jup}$, but pertinent to a model that is not 
significant. However, the same data allows us to place a much tighter
companion mass upper limit of 
$112~~\mathrm{M}_\mathrm{Jup}$ at $99.73\% $ confidence.
This mass upper limit still allows a stellar companion, but with a 
low probability. From the astrometric analysis the chance probability 
that the companion mass exceeds the stellar mass threshold is 
$2.9\% $.

If the brown dwarf nature of this object can be fully established,
e.g.~from future astrometric measurements, it would be the first
genuine brown dwarf desert object orbiting an early-M dwarf.


\begin{acknowledgements}
We thank the ESO OPC for generous allocation of observing time and
the Science Operations Team of Paranal Observatory for carrying out 
the service mode observations for this programme.
ME acknowledges support by the National Aeronautics and
Space Administration under Grants NNG05G107G issued through
the Terrestrial Planet Finder Foundation Science program and Grant
NNX07AL70G issued through the Origins of Solar Systems Program.
\end{acknowledgements}

\end{document}